\documentclass{article}
\usepackage{graphicx}
\usepackage{amssymb}
\usepackage{times}
\usepackage{emulateapj}

\def\ref{\par\noindent\hang}

\def\spose#1{\hbox to 0pt{#1\hss}}

\let\approxgt=\gtrsim

\def\multleft#1{\hbox to size{\vbox {\halign {\lft{##}\cr #1}}\hfill}\par}
\def\multright#1{\hbox to size{\vbox {\halign {\rt{##}\cr #1}}\hfill}\par}

\def\degmark{^\circ}
\def\boxit#1{\vbox{\hrule\hbox{\vrule\kern3pt\vbox{\kern3pt
          #1 \kern3pt}\kern3pt\vrule}\hrule}}

\def\cm{{\rm\thinspace cm}}

\def\erg{{\rm\thinspace erg}}

\def\km{{\rm\thinspace km}}
\def\kpc{{\rm\thinspace kpc}}

\def\s{{\rm\thinspace s}}

\def\pcmcu{\hbox{$\cm^{-3}\,$}}

\def\ergps{\hbox{$\erg\s^{-1}\,$}}

\def\kmps{\hbox{$\km\s^{-1}\,$}}

\makeatletter
\@ifundefined{chapter}{\def\thebibliography#1{
\section*{References}
  \list
  {\relax}{\setlength{\labelsep}{0em}
        \setlength{\itemindent}{-\bibhang}
        \setlength{\itemsep}{\parskip}
        \setlength{\parsep}{0pt}
        \setlength{\leftmargin}{\bibhang}}
    \def\newblock{\hskip .11em plus .33em minus .07em}
    \sloppy\clubpenalty4000\widowpenalty4000
    \sfcode`\.=1000\relax}}%
{\def\thebibliography#1{
  \list
  {\relax}{\setlength{\labelsep}{0em}
        \setlength{\itemindent}{-\bibhang}
        \setlength{\itemsep}{\parskip}
        \setlength{\parsep}{0pt}
        \setlength{\leftmargin}{\bibhang}}
    \def\newblock{\hskip .11em plus .33em minus .07em}
    \sloppy\clubpenalty4000\widowpenalty4000
    \sfcode`\.=1000\relax}}
 
\newlength{\bibhang}
\let\@internalcite\cite
\def\cite{\@ifstar{\citey}{\citefull}}
\def\citefull{\def\astroncite##1##2{##1\ ##2}\@internalcite}
\def\citey{\def\astroncite##1##2{##1\ (##2)}\@internalcite}
\def\citeyear{\def\astroncite##1##2{##2}\@internalcite}
\def\citename{\def\astroncite##1##2{##1}\@internalcite}
\def\@citex[#1]#2{\if@filesw\immediate\write\@auxout{\string\citation{#2}}\fi
  \def\@citea{}\@cite{\@for\@citeb:=#2\do
    {\@citea\def\@citea{; }\@ifundefined
       {b@\@citeb}{{\bf ??}\@warning
       {Citation `\@citeb' on page \thepage \space undefined}}%
{\csname b@\@citeb\endcsname}}}{#1}}
 
\def\@cite#1#2{#1\if@tempswa #2\fi} 
\def\@biblabel#1{}
 
\def\astroncite#1#2{#1\ #2}
\makeatother

\begin{document}

\title{Shocks and sonic booms in the intracluster medium:\\
X-ray shells and radio galaxy activity}

\author{Christopher~S.~Reynolds\altaffilmark{1,2}, Sebastian Heinz\altaffilmark{1,3}, and Mitchell C. Begelman\altaffilmark{1,3}
}

\altaffiltext{1}{JILA, Campus Box 440, University of Colorado,
Boulder CO~80309-0440}

\altaffiltext{2}{Hubble Fellow}

\altaffiltext{3}{Dept. of Astrophysical and Planetary Sciences,
University of Colorado, Boulder CO~80309-0391}

\begin{abstract} Motivated by hydrodynamic simulations, we discuss the
X-ray appearance of radio galaxies embedded in the intracluster medium
(ICM) of a galaxy cluster.  We distinguish three regimes.  In the early
life of a powerful source, the entire radio cocoon is expanding
supersonically and hence drives a strong shock into the ICM.  Eventually,
the sides of the cocoon become subsonic and the ICM is disturbed by the
sonic booms of the jet's working surface.  In both of these regimes, X-ray
observations would find an X-ray shell.  In the strong shock regime, this
shell will be hot and relatively thin.  However, in the weak shock
(sonic-boom) regime, the shell will be approximately the same temperature
as the undisturbed ICM.  If a cooling flow is present, the observed shell
may even be cooler than the undisturbed ICM due to the lifting of cooler
material into the shell from the inner (cooler) regions of the cluster. In
the third and final regime, the cocoon has collapsed and no well-defined
X-ray shell will be seen.  We discuss ways of estimating the power and age
of the source once its regime of behavior has been determined. 
\end{abstract}


\section{Introduction}

In the powerful Faranoff-Riley type II radio galaxies (FR-II; Fanaroff
\& Riley 1974), jets from a central AGN propagate at relativistic
speeds before terminating in a series of shocks resulting from
interaction with the surrounding material.  The spent jet material is
then thought to inflate a broad cocoon which encases the jets and is
seen as the observed radio lobes.  In very powerful or young sources,
the cocoon is probably overpressured (Begelman \& Cioffi 1989) with
respect to the surrounding ambient material (either the interstellar
medium [ISM] of the host galaxy, or the intracluster medium [ICM] of the
host cluster).  The cocoon then undergoes supersonic pressure driven
expansion into the ambient medium, sweeping the ambient medium into a
shocked shell.  A contact discontinuity separates the relativistic
cocoon material from the shocked ambient material.

X-ray observations provide one with a direct probe of the interaction of
the ISM/ICM with the radio galaxy cocoon.  X-ray imaging observations
with the high resolution imagers (HRIs) on the {\it ROSAT} satellite
discovered several systems in which the observable radio lobes are seen
to reside in evacuated cavities within the ICM which are bounded by
shells of enhanced emission (e.g. Cygnus-A: Carilli et al. 1994;
Perseus-A: Bohringer et al. 1992; A~4059: Huang \& Sarazin \& 1998).  In
order to further explore the shell/cavity in Cygnus-A, Clarke, Harris \&
Carilli (1997) performed 3D hydrodynamic simulations of a jet
propagating into a cluster-like atmosphere.  They found that the X-ray
cavities are readily identified with the radio cocoon (whose pressure
driven expansion excludes the ICM), and that the X-ray shells can be
identified with strong shocks in the ICM.  Kaiser \& Alexander (1999)
showed how analytic self-similar hydrodynamic models can be used to
obtain similar results.  More recently, Rizza et al. (2000) performed
3-D hydrodynamic simulations in an attempt to model subtle X-ray
structure in A~133, A~2626, and A2052.

In a complementary study, Heinz, Reynolds \& Begelman (1998; hereafter
HRB98) used a simple ``bubble'' model for the cocoon/shock system
(Begelman \& Cioffi 1989; Reynolds \& Begelman 1997) to predict the
X-ray appearance of radio galaxies as a function of their age and
power.  A central assumption used in the HRB98 model is that the
presence of an X-ray shell bounding the {\it sides} of a radio cocoon
indicates a lateral shock front, which in turn implies that the sides
of the cocoon are still expanding at supersonic speeds.  This
assumption is based upon the hydrodynamical fact that, once the
sideways expansion of the cocoon becomes subsonic, the contact
discontinuity separating the cocoon and the ICM will become
susceptible to the Rayleigh-Taylor instability on timescales shorter
than the expansion timescale of the cocoon.  One might then expect the
cocoon to collapse, with the corresponding disappearance of the X-ray
cavities.

In this letter, we follow up the study of HRB98 by using
moderate-to-high resolution 2-D hydrodynamic simulations to study the
interaction of radio-galaxies with the ICM.  We find that powerful radio
galaxies in which the entire cocoon is supersonic do indeed produce a
well defined shell/cavity structure, although backflows within the
shocked ICM shell increase its thickness substantially over that
predicted by the bubble model (also see Kaiser \& Alexander 1999).
However, we also find a shell/cavity structure in sources in which the
sides of the cocoon are expanding at subsonic speeds (or even not
expanding at all).  In this case, the cocoon has not had time to
collapse, despite the fact that Rayleigh-Taylor instabilities are
gradually destroying it.  The X-ray shell corresponds to emission from a
weak shock or compression wave (i.e. a sonic boom) which is driven into
the ICM by the supersonic motion of the cocoon head.  We find that ICM
material can be lifted into this emitting shell without passing through
a strong shock, which has important implications for the temperature
structure of such X-ray shells and entropy injection into the ICM.  This
study is particularly timely given its relevance to {\it Chandra}
observations of radio galaxies embedded within galaxy clusters.

Section~2 briefly discusses the configuration and results of our
hydrodynamic simulations.  In Section~3, we compute emissivity maps
and discuss the nature of X-ray shells and cavities.  Section~4
describes how one may use X-ray data to estimate the power and age of
a given radio galaxy.  We summarize our conclusions in Section~5.

\section{Hydrodynamic simulations}

The simulations discussed in this paper were performed using the
ZEUS3D code (Stone \& Norman 1992a,1992b; Clarke, Norman \& Fielder
1994) to solve the equations of ideal hydrodynamics.  Major advantages
of this code are its flexibility in the choice of the computational
grid, and the fact that it has been extensively tested.   In all of
the simulations reported in this paper, this code is used in its
pure-hydro form (i.e., there are no magnetic fields present) and we
assume axisymmetry, thereby reducing the computational domain to two
spatial dimensions.

We model back-to-back jets originating from a source residing in the
center of a galaxy/cluster atmosphere, i.e., no assumption regarding
reflection symmetry in the plane normal to the jets was made.  The
simulations were performed in spherical polar coordinates $(r,
\theta, \phi)$, and the computational domain was the region $r\in
(0.1,10)$.  The ambient galaxy/cluster gas was assumed to be
isothermal with a (adiabatic) sound speed of unity ($c_{\rm ICM}=1$)
and a density gradient given by a $\beta$-model with $\beta=0.5$ and
core radius $r_0=2$:
\begin{equation}
\rho(r)=\frac{1}{[1+(r/r_0)^2]^{3/4}}.
\end{equation}
The gravitational potential was set so as to keep this ambient
material in hydrostatic equilibrium.  It is assumed that the
gravitational potential is dominated by background dark matter,
i.e., the self-gravity of the gas is negligible.  This reasonable
assumption prevents us from having to solve Poisson's equation as part
of the hydrodynamic simulations.  The jets themselves are initially
conical with a half opening angle of $15\degmark$, and are in pressure
equilibrium with the ambient medium.  They are given initial densities
of $\rho_{\rm jet}$ and Mach numbers (with respect to the sound speed
of the injected jet material) of ${\cal M}$.

In addition to studying X-ray shells/cavities, these simulations were
tailored to study mixing across the contact discontinuity and the
evolution of passive sources after the jets have turned off.  An array
of runs was performed to study the influence of the systems parameters
and numerical resolution.  A full discussion of all of these
hydrodynamic simulations, together with a discussion of ICM/cocoon
mixing and the evolution of passive sources, will be presented in
Reynolds, Heinz \& Begelman (2001; hereafter RHB01).  Here, we focus
on two of those simulations.  Run~1 has $\rho_{\rm jet}=0.01$ and
${\cal M}=10.5$, giving a kinetic luminosity of $L_1=24.8$.  Run~2 has
the same Mach number but a higher jet density of $\rho_{\rm jet}=0.1$,
giving a lower kinetic luminosity of $L_2=7.7$.  Both of these runs use
a $600\times 600$ grid in the $(r,\theta)$-plane, with non-uniform
grid spacing that concentrates resolution towards the cluster center
and along the two jet axes.

We can relate quantities within the code to physical quantities by
fixing the parameters of the background medium.  Suppose we set
$r_0=100\kpc$, $c_{\rm ISM}=1000\kmps$ and a central number density of
$n_0=0.01\pcmcu$ --- values representative of galaxy clusters.  Then,
one code unit of time corresponds to 50\,Myr.  The kinetic
luminosities of the jets in Run~1 and Run~2 are then $9.3\times
10^{45}\ergps$ and $2.9\times 10^{45}\ergps$, respectively.

The detailed hydrodynamics of the jet is very similar to that found
by Lind et al. (1989).  In particular, the presence of a conical shock
in the jet sprays the jet material into a wide fan as it approaches
the end of the cocoon, thereby producing a 2-d version of the
``dentist-drill'' effect (Scheuer 1974) which spreads the jet momentum
over a large working surface.  The detailed hydrodynamics of the jet
and the contact discontinuity will be discussed in RHB01.   Here we
focus on the implications of this work for X-ray shells and cavities.

\section{Shocks and sonic booms: implications for observable X-ray shells}

Figure~1 shows density maps, temperature maps and simulated {\it
Chandra} ACIS-S X-ray maps for time $t=0.1$ from Run~1, and times
$t=1.3, 2.4$ from Run~2.  The X-ray maps assume a source inclination of
$60\degmark$, and were made by integrating the total X-ray emissivity
along lines of sight through the simulated system.  The X-ray emissivity
was derived using spectra from the {\sc mekal} thermal plasma model
(Mewe, Gronenschild \& van den Oord 1985; Arnaud \& Rothenflug 1985;
Mewe, Lemen \& van den Oord 1988; Kaastra 1992) in the {\sc xspec}
software package folded through the {\it Chandra} ACIS-S response curve.

The density maps clearly show the disturbance in the ICM, and the
contact discontinuity between the disturbed ICM and the low density
cocoon.  Kelvin-Helmholtz instabilities are clearly visible along this
contact discontinuity.  In the early stages of Run~1 (e.g., left panels
of Fig.~1), the entire cocoon is expanding supersonically and a well
defined shell of shocked ICM exists around the cocoon.  As the
simulation progresses, the sideways expansion of the cocoon becomes
subsonic and eventually halts altogether.  However, the supersonic
cocoon head still drives a sideways disturbance into the ICM (in the
form of a very weak shock, or a sonic boom) which creates a thick shell
of slightly compressed ICM surrounding the radio cocoon.  After several
ICM dynamical timescales, the cocoon will collapse almost entirely,
destroying the shell like appearance of the system.  However, there will
be a significant portion of the sources lifetime in which one would
observe a clear X-ray shell which corresponds to a sonic boom rather than a
strong shock.

This distinction between the strong shock and sonic boom is important
for the following reason.  In the shock scenario, the ICM experiences
a significant jump in both density and entropy at it crosses the shock
front.  Of course, the entropy jump is due to the irreversible
thermalization of the bulk kinetic energy of the undisturbed ICM as
viewed in the frame of reference of the shock.  This leads to a rather
thin shell of disturbed ICM that is appreciably hotter than the
undisturbed ICM.  It also injects substantial entropy into the ICM
which may build up and produce observational consequences long after
the radio galaxy has died.

On the other hand, there is a negligible entropy jump associated with
the sonic boom.  Since this is more akin to a compression wave (with
a 10--20\% jump in density across the wavefront), the only heating of
the disturbed ICM is due to adiabatic compression.  In the case of a
radio galaxy expanding into an isothermal cluster atmosphere, this
will produce a thick shell of disturbed ICM.  Immediately within the
compression wave, the material is slightly hotter due to the action of
compressional heating.  However, regions of the shell are also
slightly cooler than the undisturbed ICM due to the action of
adiabatic cooling as material is adiabatically lifted into lower
pressure parts of the atmosphere.  These effects tend to cancel,
producing an observed X-ray shell that is almost the same temperature
as the undisturbed ICM.  In the case of a cooling flow (where the
temperature decreases with decreasing radius), the shell may be
significantly cooler than the undisturbed ICM due to the lifting of
cool material into the shell.  This result, which is in stark contrast
with the strong shock scenario, may be very relevant to recent {\it
Chandra} observations of Hydra-A (McNamura et al. 2000) and Perseus-A
(Fabian et al. 2000).

It is also worth noting that a backflow exists in the disturbed ICM
which significantly depletes the amount of ICM material near the head
of the cocoon.  To illustrate this point, the amount of mass contained
in the part of the disturbed ICM shell that lies within the cone
$\theta<15\degmark$ is only half of the ICM mass which was swept up in
this cone.  This, together with the fact that the shock is rather
narrow and very hot, makes the strong shock bounding the head of the
cocoon very hard to detect in X-ray maps.  Indeed, no observations to
date have detected thermal emission from this region.

\section{Further discussion}

If we wish to measure physical parameters (e.g., kinetic luminosity
and age) of a given radio galaxy, is it important to know whether it
is in the strong or weak shock regime.  Given X-ray data of sufficient
quality, the two regimes can be easily distinguished by the
temperature and morphology of the X-ray shell --- hot narrow shells
indicate strong shocks whereas thick shells with little or no
temperature difference (or even a lower temperature) indicate weak
shocks.

In the strong shock case, the detailed hydrodynamics of the shocked
shell leads to surprisingly large deviations in the thickness of the
shell as compared with the simple bubble model.   For this reason, the
diagnostic diagrams of Heinz et al. (1998) require some modification.
New and detailed diagnostic diagrams, based upon applying scaling
relations to full hydrodynamic simulations, will be presented in a
future publication.

On the other hand, there are some robust diagnostics from the X-ray band
that can be used in the weak shock case.  Since the lateral expansion
velocity of the weak shock is the ICM sound speed, the age of the source
is given by
\begin{equation}
t_{\rm RG}\approx \frac{r_{\rm shell}}{c_{\rm ICM}}
\end{equation}
where $r_{\rm shell}$ is the lateral size of the observed disturbed
shell (which is independent of the inclination angle of the source), and
$c_{\rm ICM}$ is the adiabatic sound speed of the ICM.  Once the
temperature $T$ of the ICM has been measured, the sound speed is given
by $c_{\rm ICM}^2=5kT/3\mu m_{\rm p}$ where $\mu\approx 0.5$ is the mean
molecular weight of the gas.  Assuming an orientation and source
geometry, one can estimate the volume of the X-ray cavities $V$.  The
average kinetic power of the source is then given by
\begin{equation}
L_{\rm kin}=f\frac{pV}{t_{\rm RG}}
\end{equation}
where $p$ is the pressure of the displaced ICM which can be estimated
from X-ray observations, and $f\approxgt 2$ is a factor of order unity
which includes the enthalpy term and accounts for spatial gradients in
the pressure.

\section{Conclusions}

Guided by 2-D hydrodynamic simulations, we distinguish three regimes
of radio galaxy evolution which have importance for understanding the
structure of X-ray cavities and shells.  In the early life of a
powerful radio galaxy, the entire cocoon is expanding supersonically
and a strong shock is driven into the ICM in all directions.  One
would see a relatively narrow and hot X-ray shell surrounding the
radio lobes.  However, eventually, the lateral expansion velocity of
the contact discontinuity becomes subsonic.  There can be a
substantial phase of the source's life in which the cocoon is subsonic
but has yet to collapse and is surrounded by a rather thick shell of
disturbed ICM which has suffered a weak shock or compression wave.
X-ray observations of such a system would reveal a rather thick shell
bounding a cavity.  The temperature of this shell would be very
similar to the temperature of the undisturbed ICM.  If a cooling flow
is present, this shell would contain cooler material that has been
lifted from further down in the cluster potential.  Even if the ICM is
isothermal, regions of the disturbed shell could be slightly cooler
due to the action of adiabatic expansion.  We also note that a
backflow exists in the disturbed ICM shell which significantly
depletes the shocked shell near the head of the jet and renders it
difficult to observe.  After several ICM dynamical
timescales, the cocoon will collapse and the apparent X-ray shell will
vanish.

Finally, at very late times, after the radio galaxy activity has ceased,
the relic lobes may appear as `naked' cavities within the ICM emission
(i.e. X-ray cavities that are not bounded by shells).  Such structures
have been explored in recent work by Churazov et al. (2000).

\section*{Acknowledgements}

CSR appreciates support from Hubble Fellowship grant HF-01113.01-98A.
This grant was awarded by the Space Telescope Institute, which is
operated by the Association of Universities for Research in Astronomy,
Inc., for NASA under contract NAS 5-26555.  We also appreciate support
from NASA under LTSA grant NAG5-6337, and the National Science
Foundation under grants AST-9529170 and AST-9876887.

\begin{figure*}
\caption{Density, temperature and simulated X-ray maps (top, middle and
bottom respectively) for times $t=0.10$ of Run~1 and $t=1.3,2.4$ of
Run~2 (left, middle, and right respectively).  The left panels are
$3\times 2$ code units in extent, the middle panels are $9\times 6$ code
units and the right panels are $18\times 12$ code units.  These three
cases represent the three regimes of beheviour that we have noted.}
\end{figure*}

\end{document}